\documentclass[twocolumn,superscriptaddress,linenumbers,prc,showpacs,amsmath,amssymb,letterpaper]{revtex4}
\usepackage{graphicx,color}
\usepackage{amssymb}
\usepackage{amsmath}
\usepackage{latexsym}
\usepackage{amsxtra}
\usepackage{amscd}
\usepackage{dcolumn}% Align table columns on decimal point
\usepackage{bm}% bold math
\usepackage[breaklinks]{hyperref}
\usepackage{url}
\usepackage{breakurl}

\graphicspath{{./}}
\let\oldsqrt\sqrt
\def\sqrt{\mathpalette\DHLhksqrt}
\def\DHLhksqrt#1#2{\setbox0=\hbox{$#1\oldsqrt{#2\,}$}\dimen0=\ht0
\advance\dimen0-0.2\ht0
\setbox2=\hbox{\vrule height\ht0 depth -\dimen0}%
{\box0\lower0.4pt\box2}}

\newcommand{\nuc}[2]{$^{#1}$#2}

\begin{document}

\title{Spectroscopy of $^{46}$Ar by the (t,p) two-neutron transfer reaction}

\author{K.~Nowak}
\affiliation{Physik-Department E12, Technische Universit\"at M\"unchen, D-85748 Garching, Germany}
\author{K.~Wimmer}
\affiliation{Physik-Department E12, Technische Universit\"at M\"unchen, D-85748 Garching, Germany}
\affiliation{Department of Physics, The University of Tokyo, Hongo, Bunkyo-ku, Tokyo 113-0033, Japan}
\author{S.~Hellgartner}
\affiliation{Physik-Department E12, Technische Universit\"at M\"unchen, D-85748 Garching, Germany}
\author{D.~M\"ucher}
\affiliation{Physik-Department E12, Technische Universit\"at M\"unchen, D-85748 Garching, Germany}
\author{V.~Bildstein}
\affiliation{Physik-Department E12, Technische Universit\"at M\"unchen, D-85748 Garching, Germany}
\affiliation{Department of Physics, University of Guelph, Guelph, Ontario, N1G 2W1, Canada}
\author{J.~Diriken}
\affiliation{KU Leuven, Instituut voor Kern- en Stralingsfysica, B-3001 Heverlee, Belgium}
\author{J.~Elseviers}
\affiliation{KU Leuven, Instituut voor Kern- en Stralingsfysica, B-3001 Heverlee, Belgium}
\author{L.~P.~Gaffney}
\altaffiliation[Present address: ]{School of Engineering and Computing, University of the West of Scotland, Paisley, PA1 2BE, United Kingdom}
\affiliation{Oliver Lodge Laboratory, University of Liverpool, Liverpool L69 9ZE, United Kingdom}
\author{R.~Gernh\"auser}
\affiliation{Physik-Department E12, Technische Universit\"at M\"unchen, D-85748 Garching, Germany}
\author{J.~Iwanicki}
\affiliation{Heavy Ion Laboratory, University of Warsaw, PL-02-093 Warsaw, Poland}
\author{J.~G.~Johansen}
\affiliation{Department of Physics and Astronomy, Aarhus University, DK-8000 Aarhus C, Denmark}
\author{M.~Huyse}
\affiliation{KU Leuven, Instituut voor Kern- en Stralingsfysica, B-3001 Heverlee, Belgium}
\author{J.~Konki}
\affiliation{ISOLDE, CERN, CH-1211 Geneva 23, Switzerland}
\affiliation{University of Jyvaskyla, Department of Physics, P.O. Box 35, FI-40014, University of Jyvaskyla, Finland\\
Helsinki Institute of Physics, P.O.Box 64, FI-00014 University of Helsinki, Finland}
\author{T.~Kr\"oll}
\affiliation{Institut f\"ur Kernphysik, Technische Universit\"at Darmstadt, D-64289 Darmstadt, Germany}
\author{R.~Kr\"ucken}
\affiliation{Physik-Department E12, Technische Universit\"at M\"unchen, D-85748 Garching, Germany}
\affiliation{Department of Physics and Astronomy, University of British Columbia, Vancouver, British Columbia, Canada V6T 1Z1\\
TRIUMF, 4004 Wesbrook Mall, Vancouver, British Columbia, Canada V6T 2A3}
\author{R.~Lutter}
\affiliation{Ludwig-Maximilians-Universit\"at-M\"unchen, Schellingstra\ss e 4, 80799 M\"unchen, Germany}
\author{R.~Orlandi}
\affiliation{Instituto de Estructura de la Materia, IEM-CSIC, Madrid E-28006, Spain}
\author{J.~Pakarinen}
\affiliation{ISOLDE, CERN, CH-1211 Geneva 23, Switzerland}
\affiliation{University of Jyvaskyla, Department of Physics, P.O. Box 35, FI-40014, University of Jyvaskyla, Finland\\
Helsinki Institute of Physics, P.O.Box 64, FI-00014 University of Helsinki, Finland}
\author{R.~Raabe}
\affiliation{KU Leuven, Instituut voor Kern- en Stralingsfysica, B-3001 Heverlee, Belgium}
\author{P.~Reiter}
\affiliation{Institut f\"ur Kernphysik, Universit\"at zu K\"oln, D-50937 K\"oln, Germany}
\author{T.~Roger}
\affiliation{KU Leuven, Instituut voor Kern- en Stralingsfysica, B-3001 Heverlee, Belgium}
\author{G.~Schrieder}
\affiliation{Institut f\"ur Kernphysik, Technische Universit\"at Darmstadt, D-64289 Darmstadt, Germany}
\author{M.~Seidlitz}
\affiliation{Institut f\"ur Kernphysik, Universit\"at zu K\"oln, D-50937 K\"oln, Germany}
\author{O.~Sorlin}
\affiliation{Grand Acc\'el\'erateur National d'Ions Lourds (GANIL), CEA/DSM - CNRS/IN2P3, B.\ P.\ 55027, F-14076 Caen Cedex 5, France}
\author{P.~Van~Duppen}
\affiliation{KU Leuven, Instituut voor Kern- en Stralingsfysica, B-3001 Heverlee, Belgium}
\author{N.~Warr}
\affiliation{Institut f\"ur Kernphysik, Universit\"at zu K\"oln, D-50937 K\"oln, Germany}
\author{H.~De~Witte}
\affiliation{KU Leuven, Instituut voor Kern- en Stralingsfysica, B-3001 Heverlee, Belgium}
\author{M.~Zieli\'nska}
\affiliation{Heavy Ion Laboratory, University of Warsaw, PL-02-093 Warsaw, Poland}

\begin{abstract}
States in the $N=28$ nucleus \nuc{46}Ar have been studied by a two-neutron transfer reaction at REX-ISOLDE (CERN). A beam of radioactive \nuc{44}{Ar} at an energy of 2.16~AMeV and a tritium loaded titanium target were used to populate \nuc{46}{Ar} by the t(\nuc{44}{Ar},p) two-neutron transfer reaction. Protons emitted from the target were identified in the T-REX silicon detector array. The excitation energies of states in \nuc{46}Ar have been reconstructed from the measured angles and energies of recoil protons. Angular distributions for three final states were measured and based on the shape of the differential cross section an excited state at 3695~keV has been identified as $J^\pi = 0^+$. The angular differential cross section for the population of different states are compared to calculations using a reaction model employing both sequential and direct transfer of two neutrons. Results are compared to shell model calculations using state-of-the-art effective interactions.

\end{abstract}

\date{\today}
\maketitle

\section{Introduction}
Among the magic numbers which describe the shell structure of atomic nuclei, $28$ is the first main shell gap created by the spin-orbit interaction. The $1f_{7/2}$ orbital gets lowered in energy compared to the $1f_{5/2}$ orbital creating this gap within the $N=3$ major oscillator shell. The evolution of the shell gap at $28$ nucleons, both as function of neutron and proton number, is influenced by the nature of the spin-orbit interaction. On the neutron-rich side of the valley of stability, it has been shown that also other terms in the nucleon interaction play a role in determining the size of the $N=28$ shell gap~\cite{sorlin13}. 
Three-body forces have been successfully employed along the Ca isotopic chain ($Z=20$) to describe of the high excitation energy of the first $2^+$ state in \nuc{48}{Ca} and the increase of the $N=28$ gap between $N=20$ and 28 microscopically~\cite{holt12}. Evolution of the gap between the neutron $sd$ shell and the $1f_{7/2}$ orbital along the $N=28$ isotones is influenced by the central and tensor interaction between protons and neutrons~\cite{otsuka10}. Below \nuc{48}{Ca} a variety of features can be seen in the low-lying excitations of the $N=28$ isotones. These arise from the subtle interplay of the forces, the breakdown of the $N=28$ shell closure, and the proton sub-shell closures at $Z=16$ (sulfur) and 14 (silicon). The $B(\text{E2};\,2_1^+\rightarrow 0_\text{gs}^+)$ value for \nuc{46}{Ar} ($Z=18$) has been measured using Coulomb excitation at intermediate energies~\cite{scheit96,gade03,calinescu14} as well as extracted from the measured lifetime~\cite{mengoni10} giving conflicting results. The value determined in the Coulomb excitation experiments ($B(\text{E2};\,2_1^+\rightarrow 0_\text{gs}^+)=39(8)$~e$^2$fm$^4$~\cite{scheit96}, $44(6)$~e$^2$fm$^4$~\cite{gade03}, and $54(5)$~e$^2$fm$^4$~\cite{calinescu14}) points to a moderate deformation and collectivity in \nuc{46}{Ar} consistent with a the expectation for a semi-magic nucleus. This is supported by time-dependent Hartree-Fock-Bogoliubov calculations~\cite{ebata15} that link the increase in collectivity with respect to \nuc{48}{Ca} to a quenching of the $N=28$ shell gap. Shell model calculations on the other hand favor the result of a larger $B(\text{E2})$ value as determined by the lifetime measurement ($B(\text{E2};\,2_1^+\rightarrow 0_\text{gs}^+)=114^{+67}_{-32}$~e$^2$fm$^4$~\cite{mengoni10}). The neutron single-particle energies of the $2p_{3/2}$, $2p_{1/2}$ and $1f_{5/2}$ orbitals in \nuc{47}{Ar} have been extracted from a (d,p) transfer reaction and compared to \nuc{49}{Ca} the $N=28$ shell gap is reduced by 330(90)~keV~\cite{gaudefroy06}. Mass measurements also show a strong gap at $N=28$~\cite{meisel15} and the separation energies are well described by calculations using the SDPF-U~\cite{nowacki09} and SDPF-MU~\cite{utsuno12} effective interactions. Below \nuc{46}{Ar} the nucleus \nuc{44}{S} exhibits a low-lying excited $0^+$ state~\cite{force10} which has been interpreted as a sign of shape-coexistence. Measurements of other low-lying states~\cite{santiago11} as well as configuration mixing calculations suggest an erosion of the $N=28$ shell closure rather than shape coexistence~\cite{rodriguez11}. \nuc{42}{Si} has a very low first excited state~\cite{bastin07} and the $R_{4/2}$ ratio indicates well developed deformation~\cite{takeuchi12}. Shell model calculations predict that this nucleus is oblate in its ground state~\cite{nowacki09, utsuno12}. 

%first experimental results now available for \nuc{40}{Mg}~\cite{crawford14}

The single-particle structure of \nuc{46}{Ar} and its neighbors has been studied in several experiments. Spectroscopic factors extracted from neutron removal reactions from \nuc{46}{Ar} to \nuc{45}{Ar} gave consistent results both in transfer~\cite{lee10} and knockout reactions~\cite{gade05}. These experiments show that the ground state of \nuc{46}{Ar} is dominated by a $f_{7/2}$ configuration. Spectroscopic factors extracted from the study of the $N=27$ isotope \nuc{45}{Ar} by a one-neutron (d,p) transfer reaction also agree with shell model results~\cite{gaudefroy08}. These results suggest that the $N=28$ shell gap is still pronounced in \nuc{46}{Ar}. Even though the first excited state in \nuc{45}{Ar} $J^\pi = 3/2^-$ is located only at 542~keV, the spectroscopic strength is larger for the second excited $3/2^-$ state at 1416~keV. The low $3/2^-_1$ state has likely a complicated structure, involving also proton excitations~\cite{gaudefroy08} and can therefore not be regarded as a sign of a reduced shell gap. 
The $B(\text{E2};\,2_1^+\rightarrow 0_\text{gs}^+)$ as determined by intermediate beam energy Coulomb excitation is rather small~\cite{scheit96,gade03,calinescu14}, a result in disagreement with the shell model calculations~\cite{nowacki09, utsuno12,kaneko11} as well as calculations using the generator coordinate method with the Gogny D1S interaction~\cite{rodriguez02}. The latter calculations predict a coexistence of spherical and deformed states at low excitation energy. The collective wave function calculated for both the $0^+_\text{gs}$ and $0^+_2$ states show a mixture of oblate and prolate components, on average this leads to a slightly oblate $0^+_\text{gs}$ and prolate $0^+_2$ at $\sim 2.75$~MeV~\cite{rodriguez02}. A relatively low-lying excited $0^+$ state is also predicted by the shell model calculations at around 3~MeV (see Fig.~\ref{fig:theolevel}). Experimentally, excited states beyond the $2^+_1$ state were observed in in-beam experiments. In a proton inelastic scattering experiment~\cite{riley05} a candidate for a $3^-$ state at 4982~keV and several unassigned states around 4~MeV were found. Candidates for $0_2^+$, $2_2^+$, and $4_1^+$ states were found in fragmentation reactions~\cite{dombradi03}. The $0_2^+$ state was located at 2710~keV and tentatively assigned only based on the observation of a 1140~keV transition in coincidence with the $2^+_1 \rightarrow 0^+_\text{gs}$ transition and the comparison to calculations. From shell model calculations in reference~\cite{dombradi03} using the interaction of reference~\cite{nummela01} the $0^+$ ground state is dominated by a $0p-0h$ configuration. The first excited $0^+$ state on the other hand has a $2p-2h$ structure with two neutrons predominantly located in the $2p_{3/2}$ orbital above $N=28$.

In this work the structure of low-lying states in \nuc{46}{Ar} was studied by a (t,p) two-neutron transfer reaction in inverse kinematics. Two-neutron transfer reactions are an excellent tool to study the nature of $0^+$ states caused by neutron excitations. The angular distribution of protons from the reaction is indicative of the transferred angular momentum of the reaction. Therefore, $0^+$ states can be identified unambiguously. Furthermore, the cross section of the two-neutron transfer reaction depends on the details of the wave functions of the states involved, allowing for precise testing of theoretical models.

%evolution of $N=28$ from \nuc{48}{Ca} to \nuc{42}{Si}\cite{bastin07}.
%\nuc{46}{Ar} is a moderately collective vibrator, \nuc{44}{S} exhibits a prolate-spherical shape coexistence, the \nuc{42}{Si} nucleus is well deformed (oblate)~\cite{takeuchi12}

%triaxiality, triple shape coexistence~\cite{santiago11}
%\cite{stroberg15}
%
%two proton sub-shell closures at $Z=16$, sulfur, and $Z=14$, silicon.
%Ar: 
%intermediate energy Coulex, transfer~\cite{lee10} removal, knockout~\cite{gade05}, lifetime~\cite{mengoni10}, transfer adding \nuc{46}{Ar}(d,p)\cite{gaudefroy06}, \nuc{44}{Ar}(d,p)\cite{gaudefroy08}

%mass~\cite{meisel15}, strong $N=28$ gap
%theory~\cite{gaudefroy10,caurier14}

%beyond $N=28$~\cite{steppenbeck15,winkler12}

% (t,p) and shape coexistence~\cite{heyde11}

\section{Experimental setup}
The experiment was performed at the REX-ISOLDE facility at CERN~\cite{vanduppen11}. Radioactive \nuc{44}{Ar} nuclei were produced by impinging the 1.4~GeV proton beam from the PS booster onto a thick uranium carbide (UC$_\text{x}$) target. In order to reduce contamination from carbon dioxide CO$_2$ at the same mass number 44 the primary target was heated before the experiment. Argon as a noble gas is volatile, emerging easily from the thick target material through a cooled transfer line to remove less volatile contaminants. A forced electron beam induced arc discharge (FEBIAD) ion source~\cite{penescu10} was used to achieve a high ionization efficiency for the $1+$ charge state of \nuc{44}{Ar}. After acceleration to 30~keV the beam is sent through the high resolution separator (HRS). The HRS provides sufficient resolution to discriminate between \nuc{44}{Ar}$^{+}$ and the remaining CO$_2^+$. Doubly charged \nuc{88}{Kr}$^{2+}$ could not fully be separated and remained in the low energy beam. After mass separation, a radio frequency quadrupole cooler and buncher were employed to improve beam emittance. Ions were then accumulated and bunched in the REX trap for 60 ms before  transportation to the electron beam ion source REX EBIS for charge breeding. For the \nuc{44}{Ar} ions a maximum in the charge state distribution at $q=+13$ was achieved in 59~ms charge breeding time. Before acceleration in the REX linear accelerator the ions are separated by their mass to charge ratio $A/q$. The charge state distribution of \nuc{88}{Kr} is sufficiently different such that an $A/q$ selection of 3.3846 provided a clean beam for the experiment. Selecting a charge state of $q=+13$ also eliminated contamination from the \nuc{22}{Ne} buffer gas used in the EBIS. The ions were accelerated by the REX LINAC consisting of a RFQ, an IH structure, three 7-gap resonators followed by a 9-gap resonator. For the present experiment the beam energy was limited to 2.16~AMeV, to avoid fusion reactions with the target carrier material, therefore the 9-gap resonator was not used.

The \nuc{44}{Ar} beam at an average intensity of $2\cdot10^5/s$ was then sent to the experimental station where it impinged on a tritiated titanium foil. The target itself is a 4.5~mm wide strip of titanium foil with a thickness of $0.5$~mg/cm$^2$. The titanium is loaded with tritium at an atomic ratio of 1.3 tritium atoms per titanium atom, corresponding to an effective tritium thickness of 36~$\mu$g/cm$^2$. The target was the same one used in Ref.~\cite{wimmer10} and the decay of the tritium had reduced the effective thickness since its production in October 2010. Light reaction partners emerging from the target were detected and identified using the T-REX silicon detector array~\cite{bildstein12}. The array consists of two boxes of 140~$\mu$m thick silicon strips detectors to measure the energy loss of light particles backed by 1~mm thick unsegmented silicon detectors for total energy measurement. In the most backward direction a double-sided annular silicon strip detector was mounted. The detectors cover 65~\% of the solid angle around the target. Recoil protons, deuterons and tritons from elastic and inelastic scattering as well as transfer reaction channels are identified by their characteristic energy loss in the thin first layer of the detector stack through the $\Delta E - E$ method. In backward direction the energy of protons is not sufficient to punch through the first layer of silicon, however, the second layer can be used to discriminate protons from electron from $\beta$-decay of beam particles accidentally stopped in the chamber. The efficiency and acceptance of the array has been modeled using a GEANT4~\cite{agostinelli03} simulation of the setup~\cite{bildstein12}. The silicon array is surrounded by the MINIBALL germanium detector array~\cite{warr13}. MINIBALL consists of 24 high purity germanium crystals, each 6-fold segmented for improved granularity, allowing for better Doppler correction of detected $\gamma$-rays. Energy and efficiency calibrations were performed using standard calibration sources.

\section{Data analysis}\label{sec:exp_anal}
Light, charged recoil particles, protons, tritons and deuterons were identified using the energy loss $\Delta E$ and total kinetic energy $E$ measurements in the two layers of the T-REX silicon detectors. For particles stopped in the $\Delta E$ layer additional kinematic cuts have been applied. In laboratory backward direction both protons and deuterons have kinetic energies below the identification threshold, they are stopped in the first layer, and therefore no particle identification is possible. However, the kinetic energy of deuterons following the (t,d) reaction is very low. Therefore, a condition on scattering angle and particle energy can be used to eliminate deuterons in the spectrum.

\begin{figure}[h]
\centering
\includegraphics[width=\columnwidth]{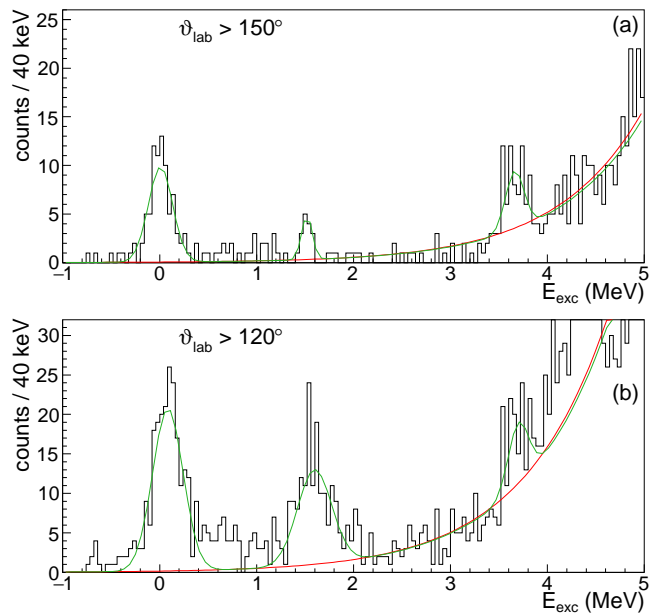}
\caption{Excitation energy of \nuc{46}{Ar} reconstructed from the proton angle and kinetic energy. The data are fit with an exponential function representing the continuum of highly excited states and Gaussian functions corresponding to states in \nuc{46}{Ar}. Panel (a) shows the most backward angles in the laboratory system, where the resolution is best. Panel (b) also includes more forward angles, where the known $2^+$ state at 1554~keV is more pronounced. Since the excitation energy resolution depends strongly on the scattering angle, the fit is only used to extract the mean position of the peaks, not the cross section.}
\label{fig:exc}
\end{figure}

The spectrum in Fig.~\ref{fig:exc} shows the excitation energy of \nuc{46}{Ar} reconstructed from the proton angle and kinetic energy.
Besides a strong population of excited states around 5~MeV (see below for details) three peaks are observed in the excitation energy spectrum. They correspond to the ground state of \nuc{46}{Ar}, the known first excited $2^+$ state at 1554~keV, and a previously unknown state at an excitation energy of 3660(60)~keV.

\begin{figure}[h]
\centering
\includegraphics[width=\columnwidth]{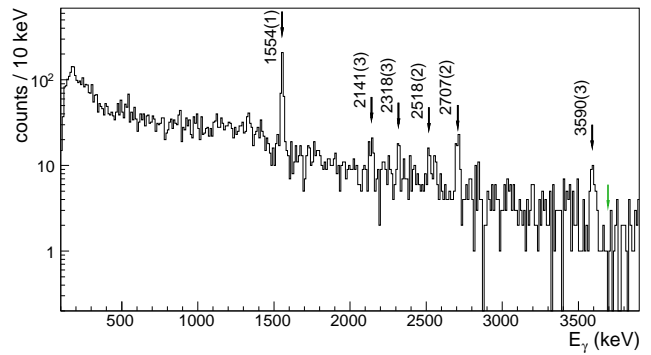}
\caption{Doppler corrected $\gamma$-ray energy spectrum measured in coincidence with recoil protons identified in T-REX. All proton angles have been included. Transitions are labeled by their energy in keV. The green arrow indicates 3695~keV, where a direct ground state decay of the proposed $0^+$ state would be located.}
\label{fig:gamma}
\end{figure}

\begin{figure}[h]
\centering
\includegraphics[width=\columnwidth]{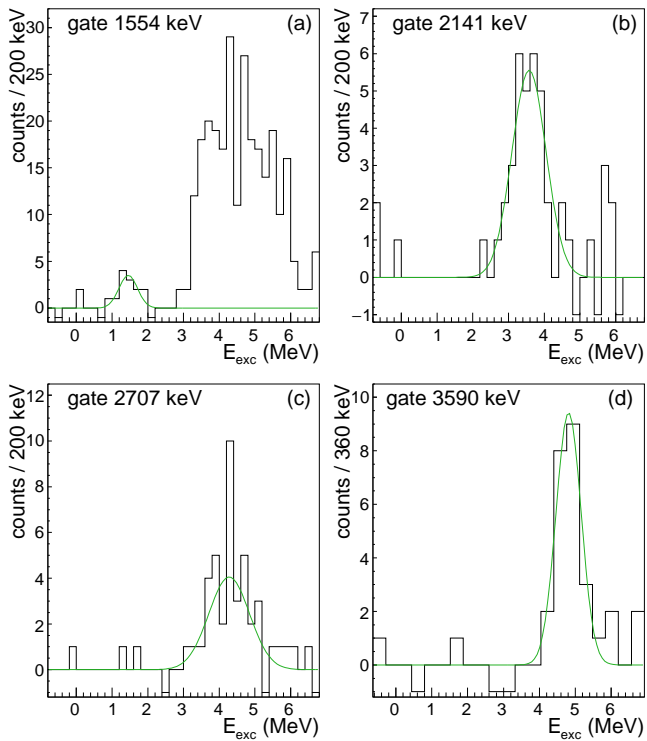}
\caption{Excitation energy of \nuc{46}{Ar} reconstructed from the proton angle and kinetic energy gated on several $\gamma$-ray transitions. All proton angles are taken into account. Random background has been subtracted. Excitation energies extracted from Gaussian fits agree with the sums of $\gamma$-ray transition energies.}
\label{fig:gamgate}
\end{figure}

\begin{figure}[h]
\centering
\includegraphics[width=0.8\columnwidth]{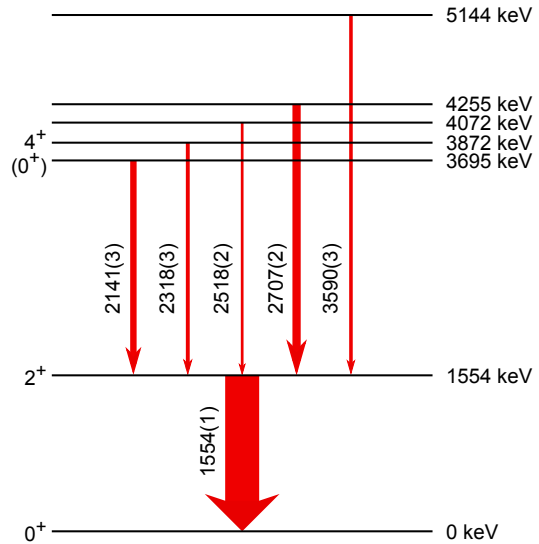}
\caption{Level scheme of \nuc{46}{Ar} as determined in this work. Arrows indicate the observed $\gamma$-ray transitions, their width corresponds to the relative intensity.}
\label{fig:level}
\end{figure}

Fig.~\ref{fig:gamma} shows the Doppler corrected $\gamma$-ray energy spectrum for \nuc{46}{Ar} assuming a scattering angle of $0^\circ$ in the laboratory system for \nuc{46}{Ar}. 
The transitions at 1554, 2318, 2518, and 2707~keV have been previously observed~\cite{dombradi03,riley05}. A transition at 1153~keV, corresponding to the decay of the previously assigned $0^+_2$ state~\cite{dombradi03} has not been observed. Newly observed are the transitions at 2141 and 3590~keV. The statistics are not sufficient for a $\gamma-\gamma$ coincidence analysis, but the analysis of the excitation energy spectrum shows that all transitions feed the first excited state and no other state below 4~MeV has been observed in Fig.~\ref{fig:exc}. Fig.~\ref{fig:gamgate} shows the excitation energy of \nuc{46}{Ar} reconstructed from the proton angle and kinetic energy measured in coincidence with the strongest $\gamma$-ray lines observed in Fig.~\ref{fig:gamma}.
The spectra have been fitted with a Gaussian function, and the resulting mean excitation energy agrees with the one determined from the sum of $\gamma$-ray energies within the error. Fig~\ref{fig:gamgate} (a) shows that the main contribution to the $2^+$ state comes from indirect feeding through excited states between 3 and 6~MeV. A gate on the 2141~keV transition reveals a single state at an excitation energy of 3670(100)~keV (Fig.~\ref{fig:gamgate} (b)). This state corresponds to the previously discussed state of Fig.~\ref{fig:exc} at 3660(60)~keV. From the sum of $\gamma$-ray transition energies the excitation energy of this state is determined to 3695(4)~keV. Similarly, we place states at 4255(4) and 5144(4)~keV which decay by 2707 and 3590~keV transition to the first excited state. For the transitions at 2318 and 2518~keV the statistics is not sufficient to determine the feeding level from proton - $\gamma$ coincidences precisely, however they arise from states around 4~MeV. These transitions are placed on top of the $2^+$ state.
The resulting level scheme of \nuc{46}{Ar} is shown in Fig.~\ref{fig:level}.
This level scheme is consistent with the one obtained from the proton inelastic scattering experiment~\cite{riley05}. Since (p,p$^\prime$) does not populate the excited $0^+$ state directly, and a two-neutron transfer reaction to a $3^-$ state is not expected, the two experiments are complementary and in good agreement.
In addition to the states shown in Fig.~\ref{fig:level} the excitation energy spectrum (Fig.~\ref{fig:exc}) indicates that several other states above 4~MeV excitation energy have been populated. The level density increases with excitation energy and many individual states are populated with small cross sections, therefore discrete lines were not identified.

\begin{figure}[h]
\centering
\includegraphics[width=\columnwidth]{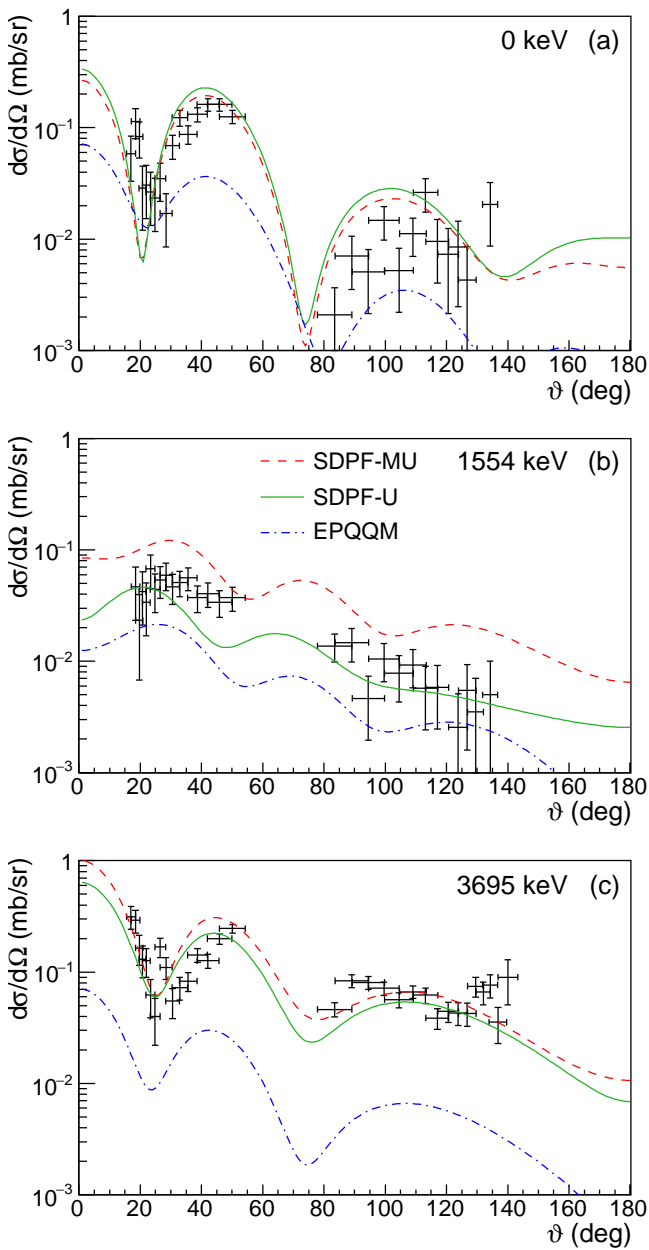}
\caption{(color online) Angular distribution of protons from the two-neutron transfer reaction to \nuc{46}{Ar}. (a) Ground state of \nuc{46}{Ar}, (b) first excited $2^+$ state, (c) excited $0^+_2$ state. Lines represent the theoretical calculations using the DWBA reaction model described in Section~\ref{sec:the_anal_reac} using amplitudes calculated with three different shell model effective interactions, SDPF-MU (red, dashed), SDPF-U (green, solid), and EPQQM (blue, dot-dashed).}
\label{fig:angdist}
\end{figure}

Since the beam intensity fluctuated during the experiment, the luminosity was determined using the elastic scattering of tritons. These data were also used to constrain the optical model parameters for the DWBA analysis. The angular distributions were obtained by gating on the excitation energy (Fig.~\ref{fig:exc}) and correcting for the geometrical acceptance of the T-REX array~\cite{bildstein12}. Fig.~\ref{fig:angdist} shows the angular distribution of protons from the two-neutron transfer reaction to the ground state and excited states of \nuc{46}{Ar} at 1554 and 3695~keV. To avoid systematic uncertainties data from the annular detector at backward angles has been excluded due to an unresolved problem with the time dependent efficiency of its multiplexed readout~\cite{bildstein12}.
The comparison with the DWBA calculations (Section~\ref{sec:the_anal_reac}) show that protons from the transfer reaction to the ground state of \nuc{46}{Ar} follow the calculated differential cross section with the characteristic $L=0$ minimum at a scattering angle $\vartheta_\text{cm}\sim20^\circ$. The angular distribution corresponding to the population of the $2^+$ state displays a shallow maximum around $\vartheta_\text{cm}\sim30^\circ$, indicative of the orbital angular momentum transfer of $L=2$. The differential cross section for the newly observed excited state at 3695~keV shows the same trend as the ground state. This characteristic $L=0$ shape as well as the $\gamma$ decay only to the $2^+$ state and not directly to the ground state, and the rather large two-neutron transfer reaction (see Section~\ref{sec:the_anal_reac}) indicate a spin and parity $J^\pi = 0^+$ for this state.

\section{Theoretical calculations}\label{sec:the_anal}
For the theoretical calculation of the two-neutron transfer reaction cross section both nuclear structure and reaction inputs are required. Shell model calculations are employed in order to obtain the spectroscopic amplitudes ($A$) for one-neutron transfer steps as well as two-nucleon amplitudes ($TNA$) for the direct pair transfer. In Section~\ref{sec:the_anal_reac} the dependence of the differential cross section on the optical model parameters and the influence of the two reaction processes are analyzed.

\subsection{Shell model calculations}\label{sec:the_anal_sm}
In order to get insights in the underlying structure causing the large cross section to the first excited $0^+$ state shell model calculations have been performed using the code NuShellX~\cite{nushell}. 
\begin{figure*}[!]
\centering
\includegraphics[width=\textwidth]{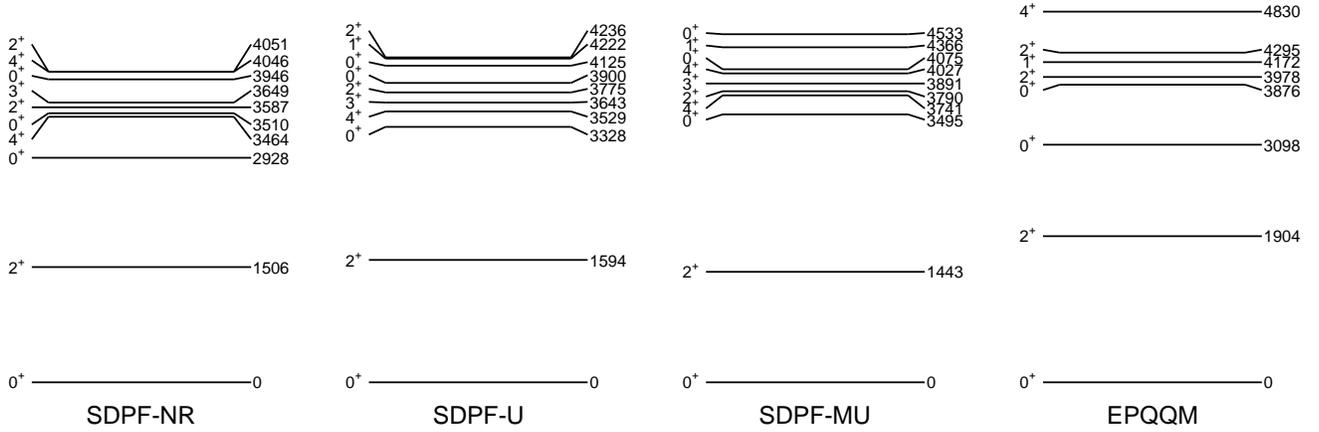}
\caption{Calculated level schemes of \nuc{46}{Ar} using various effective interactions in the shell model~\cite{nummela01,nowacki09,utsuno12,kaneko11}.}
\label{fig:theolevel}
\end{figure*}
The model space comprises of the $sd$ shell for the protons and the $fp$ shell for the neutrons. Three state-of-the-art effective interactions have been compared, SDPF-U~\cite{nowacki09}, SDPF-MU~\cite{utsuno12}, and EPQQM~\cite{kaneko11}. 
The SDPF-MU and SDPF-U interactions are constructed from three ingredients. Both use the USD~\cite{wildenthal84} effective interaction for the $sd$ proton-proton matrix elements. The neutron-neutron interaction in the $fp$ shell is based on the KB3~\cite{poves81} matrix elements for the SDPF-U interactions and the GXPF1B~\cite{honma08} interaction for SDPF-MU, respectively. The $sd-fp$ cross-shell proton-neutron matrix elements are taken from G-Matrix~\cite{kahana69} for SDPF-U interaction and from $V_{MU}$~\cite{otsuka10} in the case of SDPF-MU. The SDPF-U interaction differs from the older version, SDPF-NR~\cite{nummela01}, in that experimental information on $N=21$ and potassium nuclei has been used to constrain the monopole parts. For the present calculation the version for $Z>14$ nuclei was chosen.
The EPQQM effective interaction is based on pairing plus quadrupole-quadrupole forces with a monopole term~\cite{dufour96}. It was built in order to consistently describe nuclei between Ca and Si~\cite{kaneko11}.
The calculated level schemes of $^{46}$Ar are shown in Fig.~\ref{fig:theolevel}.
Additionally we also represent the calculations with the original SDPF-NR~\cite{nummela01} interaction, which has been previously~\cite{dombradi03} used to assign spin and parity $0^+$ to a proposed state at 2710~keV. The level scheme calculated with the SDPF-U and SDPF-MU interactions are very similar, while the EPQQM calculation predicts a higher energy for the first $2^+$ and $4^+$ states. The first excited $0^+$ state is found at lower excitation energy.

\begin{figure}[h]
\centering
\includegraphics[width=\columnwidth]{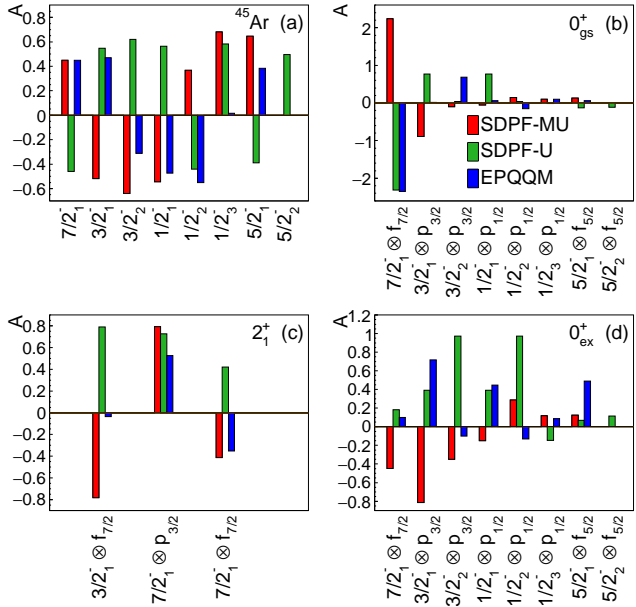}
\caption{Calculated spectroscopic amplitudes ($A$) for the transfer to states in \nuc{45}{Ar} (a), and from various states in \nuc{45}{Ar} to the ground state (b), the $2^+_1$ state (c), and the first excited $0^+$ state (d) of \nuc{46}{Ar}.}
\label{fig:sfa}
\end{figure}

\begin{figure}[h]
\centering
\includegraphics[width=\columnwidth]{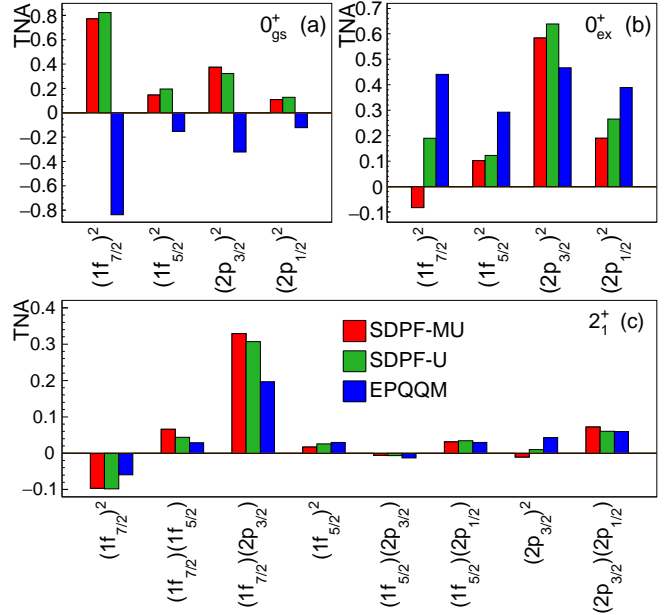}
\caption{Calculated two-nucleon amplitudes ($TNA$) for the transfer to the ground state (a), the first excited $0^+$ state (b) and the $2^+_1$ state (c).}
\label{fig:tna}
\end{figure}

The two-neutron transfer reaction can proceed either by a successive transfer of two single neutrons or by a one-step direct transfer of a neutron pair.
In order to compare the resulting two-neutron transfer cross section, the spectroscopic amplitudes $A$ for the $\langle ^{44}\text{Ar} + \text{n}|^{45}\text{Ar}\rangle$ and $\langle ^{45}\text{Ar} + \text{n}|^{46}\text{Ar}\rangle$ steps as well as two-nucleon amplitudes $TNA$ for the direct one-step transfer of a pair have been calculated. Fig.~\ref{fig:sfa} shows the spectroscopic amplitudes calculated in the shell model using the three different effective interactions.
Only states which have a calculated spectroscopic factor $C^2S = A^2 > 0.05$ are included in the figure. The cross section for a single-neutron transfer reaction such as the t(\nuc{44}{Ar},d) reaction to states in \nuc{45}{Ar} depends only on the square of the amplitude, the phase has no effect. The calculation with the SDPF-U effective interaction predicts two $5/2^-$ states with significant spectroscopic factors, which are both included in the calculation. Due to the high excitation energy, the two-step transfer reaction cross section through these states is negligible. 
For the calculation of the two-neutron transfer reaction however the relative signs matters. All the amplitudes depicted in panels (b-c), for a given effective interaction, interfere to contribute to the sequential transfer cross section. The two-nucleon amplitudes are shown in Fig.~\ref{fig:tna}.
Similar to the spectroscopic amplitudes the relative phase of the amplitudes contributing to the cross section for one state determines the interference.

\subsection{Reaction model}\label{sec:the_anal_reac}
\begin{figure}[h]
\centering
\includegraphics[width=\columnwidth]{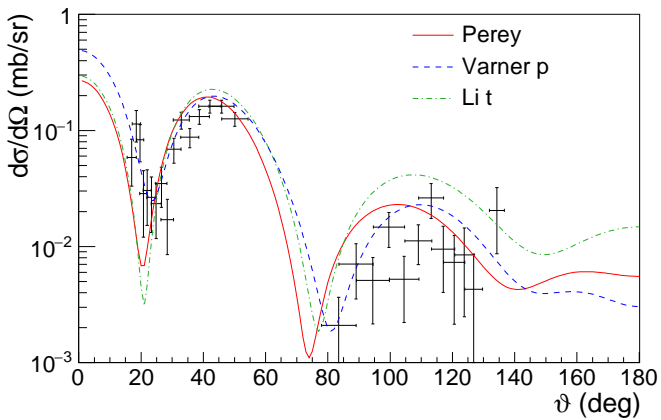}
\caption{(color online) Angular distribution of protons from the two-neutron transfer reaction to the ground state of \nuc{46}{Ar}. The solid red line represents the calculation with the parameters shown in Table~\ref{tab:opt}. For comparison we show calculations with the proton parametrization from~\cite{varner97} (blue, dashed) triton optical model parameters from~\cite{li07} (green, dot-dashed). Spectroscopic one and two-nucleon amplitudes are taken from the shell model calculation using the SDPF-MU~\cite{utsuno12} effective interaction.}
\label{fig:potentials}
\end{figure}

The two-neutron transfer reaction cross sections and angular distributions were calculated using the FRESCO DWBA code~\cite{thompson88}. Optical model parameters for the incoming, intermediate and outgoing channel were taken from global fits for tritons~\cite{perey76,li07,pang09}, deuterons~\cite{perey76,daehnick80} and protons~\cite{becchetti69,perey76,varner97}. The global parameter set of reference~\cite{perey76} is the one which is extended to the lowest projectile energies, therefore this parameter set is considered as the base line for a comparison. The numerical values of the parameters are listed in Table~\ref{tab:opt}.
\begin{table}[h]
\caption{Parameters of the optical model from reference~\cite{perey76}.}
\begin{ruledtabular}
\begin{tabular}{lrrr}
 & \nuc{44}{Ar}+t & \nuc{45}{Ar}+d & \nuc{46}{Ar}+p \\ 
\hline
$V$ (MeV) & 162.73 & 102.26 & 59.14 \\
$r$ (fm) & 1.17 & 1.05 & 1.20 \\
$a$ (fm) & 0.75 & 0.86 & 0.72 \\
$W_\text{V}$ (MeV) & 23.85 &  &  \\
$W_\text{D}$ (MeV) &  & 17.23 & 12.78 \\
$r_i$ (fm) & 1.40 & 1.43 & 1.32 \\
$a_i$ (fm) & 0.84 & 0.66 & 0.66 \\
$V_\text{SO}$ (MeV) & 2.5 & 7.0 & 6.2 \\
$r_\text{SO}$ (fm) & 1.20 & 0.75 & 1.01 \\
$a_\text{SO}$ (fm) & 0.72 & 0.50 & 0.75 \\
$r_\text{C}$ (fm) & 1.30 & 1.30 & 1.25 \\
\end{tabular}
\label{tab:opt}
\end{ruledtabular}
\end{table}
In order to estimate the effect of the potential parameters, calculations have been performed with different combinations. The result for the two-neutron transfer reaction to the ground state of \nuc{46}{Ar} is shown in Fig.~\ref{fig:potentials} for selected potentials.
With the exception of the parameter set intended for higher deuteron energies ($E_\text{d}>12$~MeV)~\cite{perey76} all parametrizations agree in their shape. The biggest impact on the shape, as well as the integrated cross section have changes in the intermediate \nuc{45}{Ar}+d channel. Since the data are not sufficient to fit the elastic scattering of tritons and protons to obtain constraints on the parameters, and for the elastic deuteron channel no data have been measured, in the following the optical potential parameters are fixed to the values listed in Table~\ref{tab:opt}. Within the angular range covered by the silicon detector array, the calculated angular distribution of elastic scattered tritons agrees with the observation.

\begin{figure}[h]
\centering
\includegraphics[width=\columnwidth]{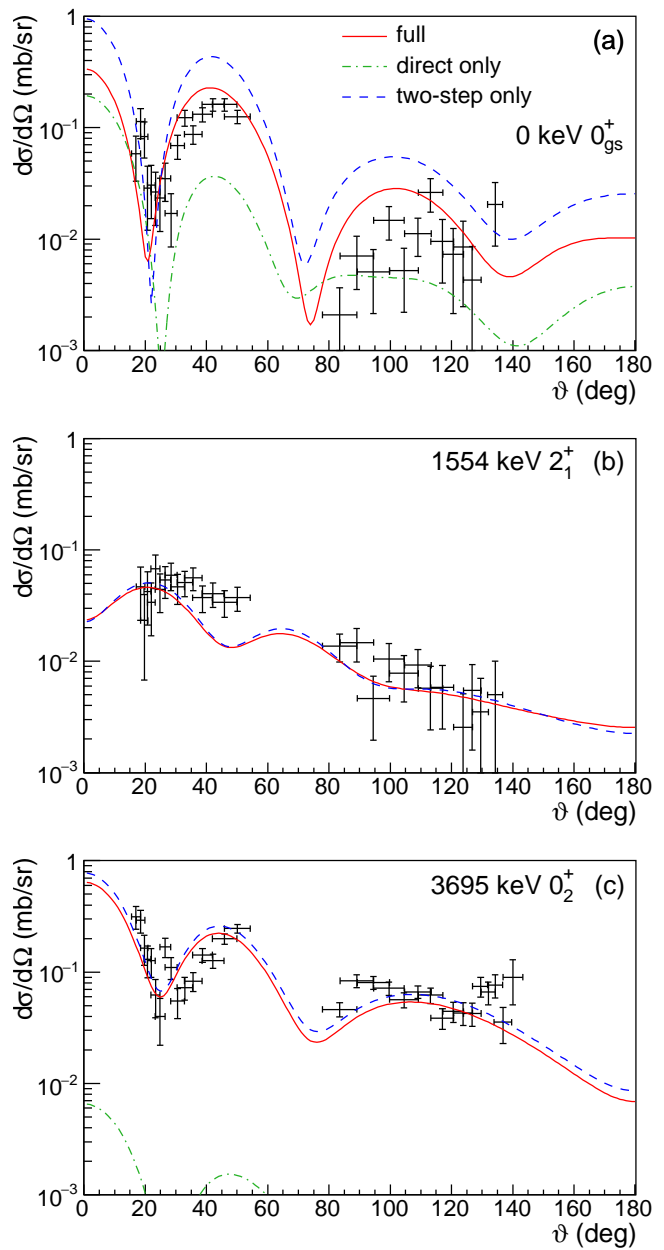}
\caption{(Color online) Calculated differential cross section for the two-neutron transfer to the ground state (a), the $2^+_1$ state (b) and the first excited $0^+$ state (c) in comparison with experimental data. The SDPF-U effective interaction has been used to calculate the one and two-nucleon amplitudes. Green dot-dashed lines represent the result including only the direct two-nucleon transfer, blue dashed lines the two-step process through states in \nuc{45}{Ar}. The solid red lines include the interference of both contributions.}
\label{fig:steps}
\end{figure}

As already indicated above, the two-neutron transfer reaction can proceed two ways, as a sequential transfer through the intermediate (\nuc{45}{Ar}+d) system or as a simultaneous direct transfer of a neutron pair. Both processes contribute to the cross section and their interference determines the total cross section. For the calculations presented in this paper the following model has been adopted.
\begin{figure}[h]
\centering
\includegraphics[width=\columnwidth]{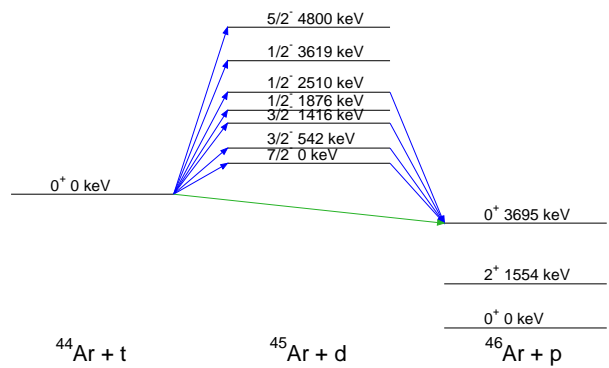}
\caption{Reaction model employed in the analysis. For the first step of the sequential (t,d)(d,p) two-neutron transfer reaction (blue arrows) several states in $^{45}$Ar have been considered as intermediate states. Experimental candidates for the theoretically calculated states with spectroscopic factors larger than 0.05 have been taken from the d($^{44}$Ar,p) measurement of reference~\cite{gaudefroy08}. The figure shows as an example the channels of the second step for which a spectroscopic factor larger than 0.05 has been calculated with the SDPF-MU interaction~\cite{utsuno12} for the excited $0^+_2$ state. See text for details.}
\label{fig:reaction}
\end{figure}
For the intermediate \nuc{45}{Ar} nucleus the ground state has been established as $J^\pi = 7/2^-$ from transfer~\cite{gaudefroy08} and knockout reactions~\cite{gade05}. The first excited state is $3/2^-$~\cite{dombradi03}. In the d(\nuc{44}Ar,p) reaction, three other $L=1$ states have been observed~\cite{gaudefroy08}. Based on the shell model calculations in Section~\ref{sec:the_anal_sm} two $3/2^-$ and three states with $J^\pi=1/2^-$ are expected to be populated strongly (Fig.~\ref{fig:sfa} (a)). The reaction model includes states with calculated spectroscopic factors larger than 0.05. The third $1/2^-$ state has no experimentally observed equivalent, therefore the excitation energy of this level is set to 3619~keV, the result of the shell model calculation using the SDPF-MU effective interaction~\cite{utsuno12}. Shell model calculations also predict a $5/2^-$ state with significant spectroscopic factor, for this the energy value for the $L=3$ candidate from transfer reactions~\cite{gaudefroy08}, 4.8~MeV, is adopted. The level lies very closely to the neutron separation energy of \nuc{45}{Ar} ($S_\text{n} = 5.169$~MeV), therefore it is suppressed by the (t,d) reaction ($Q$-value -1088~keV) and transfer through it is negligible. For the second step of the reaction, the (d,p) transfer to states in \nuc{46}{Ar} the transitions from all levels in \nuc{45}{Ar} which have a substantial spectroscopic factor calculated ($C^2S>0.05$) are included in the reaction model. Fig.~\ref{fig:reaction} shows the paths included in the calculation of the reaction to the excited $0^+_2$ state in \nuc{46}{Ar}. All spectroscopic amplitudes are implemented with their respective phase. The sequential transfer has been calculated using ``post-post'' couplings~\cite{thompson88,thompson12}, if other combinations of ``prior'' and ``post'' couplings are used, the magnitude and shape of the differential cross section varies less than if different parametrizations for the optical model are used.
For the direct one-step transfer two-nucleon amplitudes ($TNA$) are calculated. 
The results for the two components and their interference is shown in Fig.~\ref{fig:steps}.
For all three states the two-step process dominates the cross section, however the interference of one- and two-step reaction amplitudes is critical for the magnitude and shape of the differential cross section. For the ground state the direct transfer has a larger influence than for the excited $0^+_2$ state, since two-step reactions are inhibited by the reaction $Q$-value. In the case of the $2^+$ state the direct transfer alone is about two orders of magnitude smaller than the sequential one, and therefore plays a minor role. %This is due to the larger number of possible angular momentum couplings available for the ,wrong!

\section{Discussion}
While shell model predicts a larger $B(\text{E2};\,2_1^+\rightarrow 0_\text{gs}^+)$ value ($105$ e$^2$fm$^4$ calculated with the SDPF-U interaction) for \nuc{46}{Ar} than observed in Coulomb excitation, the two-neutron transfer cross section seems to be well represented. The calculation with the SDPF-U effective interaction~\cite{nowacki09} gives a better representation of the cross section to the $2^+_1$ state when standard optical model parameters are used (see Fig.~\ref{fig:angdist}). Even if different sets of parameters are used, the angular differential cross section using the amplitudes calculated with the SDPF-U effective interaction reproduces the data best.

\begin{figure}[h]
\centering
\includegraphics[width=\columnwidth]{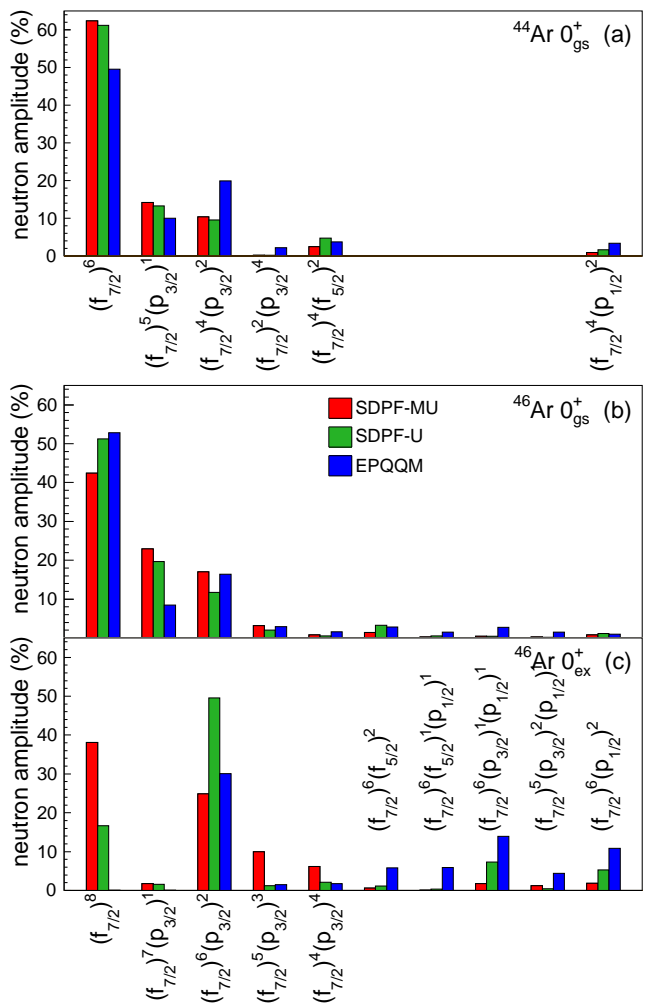}
\caption{Calculated wave functions of the ground state of \nuc{44}{Ar} (a) and two $0^+$ states in \nuc{46}{Ar}, (b) for the ground state and (c) for the first excited $0^+$ state. Only the largest neutron components are shown.}
\label{fig:wfcomp}
\end{figure}

Experimentally the cross section for the population of the ground and first excited $0^+$ states are similar in magnitude. Neutron removal reactions from the ground state of \nuc{46}{Ar}~\cite{lee10,gade05} showed that it is dominated by $0p-0h$ configurations with all valence neutrons in the $1f_{7/2}$ orbital $(f_{7/2})^8$. 
The structure of the two $0^+$ states is very different. This can be seen by looking at the contribution of different neutron particle-hole configurations to the total wave function shown in Fig.~\ref{fig:wfcomp}.
The ground state neutron configurations calculated with different effective interactions are very similar with around 50\% $0p-0h$ and around 20\% of $1p-1h$ and $2p-2h$ excitations to the $2p_{3/2}$ orbital. The configuration of the excited $0^+_2$ state is dominated by particle hole excitations. Here a striking difference between the three effective interactions can be seen. In calculations with the SDPF-MU interaction~\cite{utsuno12} $(f_{7/2})^8$ remains the largest component, $2p-2h$, $3p-3h$ and $4p-4h$ excitations to the $2p_{3/2}$ orbital contribute about 50\%. The EPQQM calculations predict a strongly mixed wave function, with many components with significant amplitudes. Interestingly, the $0p-0h$ component is absent. For the SDPF-U interaction a large component of $(f_{7/2})^6 (p_{3/2})^2$ is dominating the wave function.
The evolution in collectivity below \nuc{48}{Ca} has been attributed to the tensor component of the nuclear interaction~\cite{utsuno12}. The monopole component of the tensor interaction causes a reduction of the splitting between the $1f_{7/2}$ and $2p_{3/2}$ orbitals resulting a reduction of the $N=28$ shell-gap. In \nuc{46}{Ar} just two protons below the doubly magic \nuc{48}{Ca} the situation is unclear. The strongest among the $T=0$ cross-shell monopole terms is the attractive $\nu 1f_{7/2} - \pi 1d_{3/2}$, therefore a reduced occupation of the proton $1d_{3/2}$ orbital will cause rising of the $\nu 1f_{7/2}$ with respect to the $2p_{3/2}$ orbital compared to \nuc{48}{Ca} and reduce the $N=28$ shell gap. We have performed calculations without the cross-shell tensor force components to investigate the effect on the observables. The tensor components were removed from the SDPF-MU effective interaction. The low-lying levels of \nuc{46}{Ar} are affected by this change, the excitation energies of the $2^+_1$ and $0^+_2$ states change by only about 50 keV, however the $4^+_1$ state is lowered in energy to below the $0^+_2$ state. The effect on the two-neutron transfer cross section is more dramatic. The cross section for the (t,p) reaction to the ground state of \nuc{46}{Ar} is only slightly reduced in magnitude when the tensor components are removed from the interaction. The $2^+$ state remains unaffected by the change. Mainly the wave function composition of the excited $0^+_2$ state is altered by the removal of tensor components in the interaction. The cross section to the $0^+_2$ state is reduced by a factor of 5, in disagreement with the data. 
Even though many amplitudes contribute to the final two-neutron transfer cross section, this behavior indicates that the cross-shell proton-neutron tensor interaction has measurable effects on the observables at low excitation energy already in \nuc{46}{Ar}. 

%significantly different structure of the $0^+$ states, ground state is $(f_{7/2})^2$, excited state $(p_{3/2})^2$, EPQQM calculations differ, $0^+_2$ state is strongly mixed (see Fig.~\ref{fig:tna}), maybe calculate occupation numbers

\section{Summary and conclusions}
The $N=28$ nucleus \nuc{46}{Ar} has been studied by a (t,p) two-neutron transfer reaction at 2.16~AMeV beam energy using a radioactive tritium target. Angular distributions of protons following the population of three states are analyzed, including a previously unknown excited $0^+$ state at 3695~keV. Earlier reports of a $0^+$ state at 2710~keV~\cite{dombradi03} could not be confirmed. The differential cross sections for the population of the ground state, $2^+_1$ and $0^+_2$ states are compared to DWBA calculations including two-step reactions through the intermediate nucleus \nuc{45}{Ar} as well as the direct pair transfer. The results are robust with respect to changes in the optical model parameters for the distorted wave approximation. Spectroscopic amplitudes for the single-neutron transfer steps and two-nucleon amplitudes have been calculated in the shell model using various effective interactions. While the SDPF-MU~\cite{utsuno12} and SDPF-U~\cite{nowacki09} calculations yield comparable results for the level schemes and the cross sections to the two $0^+$ states, the cross section and energy for the $2^+_1$ state are better represented by the SDPF-U calculation. The cross section calculated with the structure input from the EPQQM~\cite{kaneko11} effective interaction is lower than the experimentally observed one for all three states. The structure of the first excited $0^+$ state differs significantly between the different interactions.%with about equal two-nucleon amplitudes for the four components.

The discrepancy between the measured and calculated $B(\text{E2};\,2_1^+\rightarrow 0_\text{gs}^+)$ value for \nuc{46}{Ar} remains unsolved. For the future we suggest to measure the $B(\text{E2})$ value as well as quadrupole moments through low-energy Coulomb excitation of \nuc{46}{Ar} like it was done for \nuc{44}{Ar}~\cite{zielinska09}.

\acknowledgments
This work has been supported by the European Commission within the Seventh Framework Programme (FP7) through ENSAR (contract number 262010), by the German BMBF (contract numbers 05P12WOFNF, 05P12PKFNE, 05P15PKCIA, 05P12RDCIA, and 05P15RDCIA), by the DFG Cluster of Excellence Origin and Structure of the Universe, by HIC for FAIR, by FWO-Vlaanderen (Belgium), by BOF KU Leuven (grant GOA/2010/010), by the Interuniversity Attraction Poles Programme initiated by the Belgian Science Policy Office (BriX network P7/12), by the UK Science and Technologies Facilities Council (STFC), and in part by the Spanish MEC Consolider - Ingenio 2010, Project No. CDS2007-00042 (CPAN).
\bibliography{draft}

\begin{thebibliography}{46}
\expandafter\ifx\csname natexlab\endcsname\relax\def\natexlab#1{#1}\fi
\expandafter\ifx\csname bibnamefont\endcsname\relax
  \def\bibnamefont#1{#1}\fi
\expandafter\ifx\csname bibfnamefont\endcsname\relax
  \def\bibfnamefont#1{#1}\fi
\expandafter\ifx\csname citenamefont\endcsname\relax
  \def\citenamefont#1{#1}\fi
\expandafter\ifx\csname url\endcsname\relax
  \def\url#1{\texttt{#1}}\fi
\expandafter\ifx\csname urlprefix\endcsname\relax\def\urlprefix{URL }\fi
\providecommand{\bibinfo}[2]{#2}
\providecommand{\eprint}[2][]{\url{#2}}

\bibitem[{\citenamefont{Sorlin and Porquet}(2013)}]{sorlin13}
\bibinfo{author}{\bibfnamefont{O.}~\bibnamefont{Sorlin}} \bibnamefont{and}
  \bibinfo{author}{\bibfnamefont{M.-G.} \bibnamefont{Porquet}},
  \bibinfo{journal}{Phys. Scr. T} \textbf{\bibinfo{volume}{12}},
  \bibinfo{pages}{014003} (\bibinfo{year}{2013}).

\bibitem[{\citenamefont{Holt et~al.}(2012)}]{holt12}
\bibinfo{author}{\bibfnamefont{J.~D.} \bibnamefont{Holt}} \bibnamefont{et~al.},
  \bibinfo{journal}{Journal of Physics G} \textbf{\bibinfo{volume}{39}},
  \bibinfo{pages}{085111} (\bibinfo{year}{2012}).

\bibitem[{\citenamefont{Otsuka et~al.}(2010)}]{otsuka10}
\bibinfo{author}{\bibfnamefont{T.}~\bibnamefont{Otsuka}} \bibnamefont{et~al.},
  \bibinfo{journal}{Phys. Rev. Lett.} \textbf{\bibinfo{volume}{104}},
  \bibinfo{pages}{012501} (\bibinfo{year}{2010}).

\bibitem[{\citenamefont{Scheit et~al.}(1996)}]{scheit96}
\bibinfo{author}{\bibfnamefont{H.}~\bibnamefont{Scheit}} \bibnamefont{et~al.},
  \bibinfo{journal}{Phys. Rev. Lett.} \textbf{\bibinfo{volume}{77}},
  \bibinfo{pages}{3967} (\bibinfo{year}{1996}).

\bibitem[{\citenamefont{Gade et~al.}(2003)}]{gade03}
\bibinfo{author}{\bibfnamefont{A.}~\bibnamefont{Gade}} \bibnamefont{et~al.},
  \bibinfo{journal}{Phys. Rev. C} \textbf{\bibinfo{volume}{68}},
  \bibinfo{pages}{014302} (\bibinfo{year}{2003}).

\bibitem[{\citenamefont{Calinescu et~al.}(2014)}]{calinescu14}
\bibinfo{author}{\bibfnamefont{S.}~\bibnamefont{Calinescu}}
  \bibnamefont{et~al.}, \bibinfo{journal}{Acta Phys. Pol. B}
  \textbf{\bibinfo{volume}{45}}, \bibinfo{pages}{199} (\bibinfo{year}{2014}).

\bibitem[{\citenamefont{Mengoni et~al.}(2010)}]{mengoni10}
\bibinfo{author}{\bibfnamefont{D.}~\bibnamefont{Mengoni}} \bibnamefont{et~al.},
  \bibinfo{journal}{Phys. Rev. C} \textbf{\bibinfo{volume}{82}},
  \bibinfo{pages}{024308} (\bibinfo{year}{2010}).

\bibitem[{\citenamefont{Ebata and Kimura}(2015)}]{ebata15}
\bibinfo{author}{\bibfnamefont{S.}~\bibnamefont{Ebata}} \bibnamefont{and}
  \bibinfo{author}{\bibfnamefont{M.}~\bibnamefont{Kimura}},
  \bibinfo{journal}{Phys. Rev. C} \textbf{\bibinfo{volume}{91}},
  \bibinfo{pages}{014309} (\bibinfo{year}{2015}).

\bibitem[{\citenamefont{Gaudefroy et~al.}(2006)}]{gaudefroy06}
\bibinfo{author}{\bibfnamefont{L.}~\bibnamefont{Gaudefroy}}
  \bibnamefont{et~al.}, \bibinfo{journal}{Phys. Rev. Lett.}
  \textbf{\bibinfo{volume}{97}}, \bibinfo{pages}{092501}
  (\bibinfo{year}{2006}).

\bibitem[{\citenamefont{Meisel et~al.}(2015)}]{meisel15}
\bibinfo{author}{\bibfnamefont{Z.}~\bibnamefont{Meisel}} \bibnamefont{et~al.},
  \bibinfo{journal}{Phys. Rev. Lett.} \textbf{\bibinfo{volume}{114}},
  \bibinfo{pages}{022501} (\bibinfo{year}{2015}).

\bibitem[{\citenamefont{Nowacki and Poves}(2009)}]{nowacki09}
\bibinfo{author}{\bibfnamefont{F.}~\bibnamefont{Nowacki}} \bibnamefont{and}
  \bibinfo{author}{\bibfnamefont{A.}~\bibnamefont{Poves}},
  \bibinfo{journal}{Phys. Rev. C} \textbf{\bibinfo{volume}{79}},
  \bibinfo{pages}{014310} (\bibinfo{year}{2009}).

\bibitem[{\citenamefont{Utsuno et~al.}(2012)\citenamefont{Utsuno, Otsuka,
  Brown, Honma, Mizusaki, and Shimizu}}]{utsuno12}
\bibinfo{author}{\bibfnamefont{Y.}~\bibnamefont{Utsuno}},
  \bibinfo{author}{\bibfnamefont{T.}~\bibnamefont{Otsuka}},
  \bibinfo{author}{\bibfnamefont{B.~A.} \bibnamefont{Brown}},
  \bibinfo{author}{\bibfnamefont{M.}~\bibnamefont{Honma}},
  \bibinfo{author}{\bibfnamefont{T.}~\bibnamefont{Mizusaki}}, \bibnamefont{and}
  \bibinfo{author}{\bibfnamefont{N.}~\bibnamefont{Shimizu}},
  \bibinfo{journal}{Phys. Rev. C} \textbf{\bibinfo{volume}{86}},
  \bibinfo{pages}{051301} (\bibinfo{year}{2012}).

\bibitem[{\citenamefont{Force et~al.}(2010)}]{force10}
\bibinfo{author}{\bibfnamefont{C.}~\bibnamefont{Force}} \bibnamefont{et~al.},
  \bibinfo{journal}{Phys. Rev. Lett.} \textbf{\bibinfo{volume}{105}},
  \bibinfo{pages}{102501} (\bibinfo{year}{2010}).

\bibitem[{\citenamefont{Santiago-Gonzalez et~al.}(2011)}]{santiago11}
\bibinfo{author}{\bibfnamefont{D.}~\bibnamefont{Santiago-Gonzalez}}
  \bibnamefont{et~al.}, \bibinfo{journal}{Phys. Rev. C}
  \textbf{\bibinfo{volume}{83}}, \bibinfo{pages}{061305}
  (\bibinfo{year}{2011}).

\bibitem[{\citenamefont{Rodr\'{\i}guez and Egido}(2011)}]{rodriguez11}
\bibinfo{author}{\bibfnamefont{T.~R.} \bibnamefont{Rodr\'{\i}guez}}
  \bibnamefont{and} \bibinfo{author}{\bibfnamefont{J.~L.} \bibnamefont{Egido}},
  \bibinfo{journal}{Phys. Rev. C} \textbf{\bibinfo{volume}{84}},
  \bibinfo{pages}{051307} (\bibinfo{year}{2011}).

\bibitem[{\citenamefont{Bastin et~al.}(2007)}]{bastin07}
\bibinfo{author}{\bibfnamefont{B.}~\bibnamefont{Bastin}} \bibnamefont{et~al.},
  \bibinfo{journal}{Phys. Rev. Lett.} \textbf{\bibinfo{volume}{99}},
  \bibinfo{pages}{022503} (\bibinfo{year}{2007}).

\bibitem[{\citenamefont{Takeuchi et~al.}(2012)}]{takeuchi12}
\bibinfo{author}{\bibfnamefont{S.}~\bibnamefont{Takeuchi}}
  \bibnamefont{et~al.}, \bibinfo{journal}{Phys. Rev. Lett.}
  \textbf{\bibinfo{volume}{109}}, \bibinfo{pages}{182501}
  (\bibinfo{year}{2012}).

\bibitem[{\citenamefont{Lee et~al.}(2010)\citenamefont{Lee, Tsang, Bazin
  et~al.}}]{lee10}
\bibinfo{author}{\bibfnamefont{J.}~\bibnamefont{Lee}},
  \bibinfo{author}{\bibfnamefont{M.~B.} \bibnamefont{Tsang}},
  \bibinfo{author}{\bibfnamefont{D.}~\bibnamefont{Bazin}},
  \bibnamefont{et~al.}, \bibinfo{journal}{Phys. Rev. Lett.}
  \textbf{\bibinfo{volume}{104}}, \bibinfo{pages}{112701}
  (\bibinfo{year}{2010}).

\bibitem[{\citenamefont{Gade et~al.}(2005)\citenamefont{Gade, Bazin, Bertulani
  et~al.}}]{gade05}
\bibinfo{author}{\bibfnamefont{A.}~\bibnamefont{Gade}},
  \bibinfo{author}{\bibfnamefont{D.}~\bibnamefont{Bazin}},
  \bibinfo{author}{\bibfnamefont{C.~A.} \bibnamefont{Bertulani}},
  \bibnamefont{et~al.}, \bibinfo{journal}{Phys. Rev. C}
  \textbf{\bibinfo{volume}{71}}, \bibinfo{pages}{051301}
  (\bibinfo{year}{2005}).

\bibitem[{\citenamefont{Gaudefroy et~al.}(2008)}]{gaudefroy08}
\bibinfo{author}{\bibfnamefont{L.}~\bibnamefont{Gaudefroy}}
  \bibnamefont{et~al.}, \bibinfo{journal}{Phys. Rev. C}
  \textbf{\bibinfo{volume}{78}}, \bibinfo{pages}{034307}
  (\bibinfo{year}{2008}).

\bibitem[{\citenamefont{Kaneko et~al.}(2011)\citenamefont{Kaneko, Sun,
  Mizusaki, and Hasegawa}}]{kaneko11}
\bibinfo{author}{\bibfnamefont{K.}~\bibnamefont{Kaneko}},
  \bibinfo{author}{\bibfnamefont{Y.}~\bibnamefont{Sun}},
  \bibinfo{author}{\bibfnamefont{T.}~\bibnamefont{Mizusaki}}, \bibnamefont{and}
  \bibinfo{author}{\bibfnamefont{M.}~\bibnamefont{Hasegawa}},
  \bibinfo{journal}{Phys. Rev. C} \textbf{\bibinfo{volume}{83}},
  \bibinfo{pages}{014320} (\bibinfo{year}{2011}).

\bibitem[{\citenamefont{Rodr\'{\i}guez-Guzm\'an
  et~al.}(2002)\citenamefont{Rodr\'{\i}guez-Guzm\'an, Egido, and
  Robledo}}]{rodriguez02}
\bibinfo{author}{\bibfnamefont{R.}~\bibnamefont{Rodr\'{\i}guez-Guzm\'an}},
  \bibinfo{author}{\bibfnamefont{J.~L.} \bibnamefont{Egido}}, \bibnamefont{and}
  \bibinfo{author}{\bibfnamefont{L.~M.} \bibnamefont{Robledo}},
  \bibinfo{journal}{Phys. Rev. C} \textbf{\bibinfo{volume}{65}},
  \bibinfo{pages}{024304} (\bibinfo{year}{2002}).

\bibitem[{\citenamefont{Riley et~al.}(2005)}]{riley05}
\bibinfo{author}{\bibfnamefont{L.~A.} \bibnamefont{Riley}}
  \bibnamefont{et~al.}, \bibinfo{journal}{Phys. Rev. C}
  \textbf{\bibinfo{volume}{72}}, \bibinfo{pages}{024311}
  (\bibinfo{year}{2005}).

\bibitem[{\citenamefont{Dombradi et~al.}(2003)}]{dombradi03}
\bibinfo{author}{\bibfnamefont{Z.}~\bibnamefont{Dombradi}}
  \bibnamefont{et~al.}, \bibinfo{journal}{Nucl. Phys. A}
  \textbf{\bibinfo{volume}{727}}, \bibinfo{pages}{195} (\bibinfo{year}{2003}).

\bibitem[{\citenamefont{Nummela et~al.}(2001)}]{nummela01}
\bibinfo{author}{\bibfnamefont{S.}~\bibnamefont{Nummela}} \bibnamefont{et~al.},
  \bibinfo{journal}{Phys. Rev. C} \textbf{\bibinfo{volume}{63}},
  \bibinfo{pages}{044316} (\bibinfo{year}{2001}).

\bibitem[{\citenamefont{Duppen and Riisager}(2011)}]{vanduppen11}
\bibinfo{author}{\bibfnamefont{P.~V.} \bibnamefont{Duppen}} \bibnamefont{and}
  \bibinfo{author}{\bibfnamefont{K.}~\bibnamefont{Riisager}},
  \bibinfo{journal}{Journal of Physics G: Nuclear and Particle Physics}
  \textbf{\bibinfo{volume}{38}}, \bibinfo{pages}{024005}
  (\bibinfo{year}{2011}).

\bibitem[{\citenamefont{Penescu et~al.}(2010)\citenamefont{Penescu, Catherall,
  Lettry, and Stora}}]{penescu10}
\bibinfo{author}{\bibfnamefont{L.}~\bibnamefont{Penescu}},
  \bibinfo{author}{\bibfnamefont{R.}~\bibnamefont{Catherall}},
  \bibinfo{author}{\bibfnamefont{J.}~\bibnamefont{Lettry}}, \bibnamefont{and}
  \bibinfo{author}{\bibfnamefont{T.}~\bibnamefont{Stora}},
  \bibinfo{journal}{Rev. Sci. Instr.} \textbf{\bibinfo{volume}{81}},
  \bibinfo{pages}{02A906} (\bibinfo{year}{2010}).

\bibitem[{\citenamefont{Wimmer et~al.}(2010)\citenamefont{Wimmer, Kr\"oll,
  Kr\"ucken et~al.}}]{wimmer10}
\bibinfo{author}{\bibfnamefont{K.}~\bibnamefont{Wimmer}},
  \bibinfo{author}{\bibfnamefont{T.}~\bibnamefont{Kr\"oll}},
  \bibinfo{author}{\bibfnamefont{R.}~\bibnamefont{Kr\"ucken}},
  \bibnamefont{et~al.}, \bibinfo{journal}{Phys. Rev. Lett.}
  \textbf{\bibinfo{volume}{105}}, \bibinfo{pages}{252501}
  (\bibinfo{year}{2010}).

\bibitem[{\citenamefont{Bildstein et~al.}(2012)\citenamefont{Bildstein,
  Gernh{\"a}user, T.Kr{\"o}ll et~al.}}]{bildstein12}
\bibinfo{author}{\bibfnamefont{V.}~\bibnamefont{Bildstein}},
  \bibinfo{author}{\bibfnamefont{R.}~\bibnamefont{Gernh{\"a}user}},
  \bibinfo{author}{\bibnamefont{T.Kr{\"o}ll}}, \bibnamefont{et~al.},
  \bibinfo{journal}{The European Physical Journal A}
  \textbf{\bibinfo{volume}{48}}, \bibinfo{pages}{85} (\bibinfo{year}{2012}).

\bibitem[{\citenamefont{Agostinelli et~al.}(2003)\citenamefont{Agostinelli,
  Allison, Amako, Apostolakis et~al.}}]{agostinelli03}
\bibinfo{author}{\bibfnamefont{S.}~\bibnamefont{Agostinelli}},
  \bibinfo{author}{\bibfnamefont{J.}~\bibnamefont{Allison}},
  \bibinfo{author}{\bibfnamefont{K.}~\bibnamefont{Amako}},
  \bibinfo{author}{\bibfnamefont{J.}~\bibnamefont{Apostolakis}},
  \bibnamefont{et~al.}, \bibinfo{journal}{Nucl. Instrum. Methods Phys. Res. A}
  \textbf{\bibinfo{volume}{506}}, \bibinfo{pages}{250} (\bibinfo{year}{2003}).

\bibitem[{\citenamefont{Warr et~al.}(2013)\citenamefont{Warr, Van~de Walle
  et~al.}}]{warr13}
\bibinfo{author}{\bibfnamefont{N.}~\bibnamefont{Warr}},
  \bibinfo{author}{\bibfnamefont{J.}~\bibnamefont{Van~de Walle}},
  \bibnamefont{et~al.}, \bibinfo{journal}{The European Physical Journal A}
  \textbf{\bibinfo{volume}{49}}, \bibinfo{pages}{40} (\bibinfo{year}{2013}).

\bibitem[{\citenamefont{Brown et~al.}()}]{nushell}
\bibinfo{author}{\bibfnamefont{B.~A.} \bibnamefont{Brown}}
  \bibnamefont{et~al.}, \emph{\bibinfo{title}{{N}u{S}hell{X}}},
  \urlprefix\url{https://people.nscl.msu.edu/~brown/resources/resources.html}.

\bibitem[{\citenamefont{Wildenthal}(1984)}]{wildenthal84}
\bibinfo{author}{\bibfnamefont{B.}~\bibnamefont{Wildenthal}},
  \bibinfo{journal}{Progress in Particle and Nuclear Physics}
  \textbf{\bibinfo{volume}{11}}, \bibinfo{pages}{5} (\bibinfo{year}{1984}).

\bibitem[{\citenamefont{Poves and Zuker}(1981)}]{poves81}
\bibinfo{author}{\bibfnamefont{A.}~\bibnamefont{Poves}} \bibnamefont{and}
  \bibinfo{author}{\bibfnamefont{A.}~\bibnamefont{Zuker}},
  \bibinfo{journal}{Physics Reports} \textbf{\bibinfo{volume}{70}},
  \bibinfo{pages}{235 } (\bibinfo{year}{1981}).

\bibitem[{\citenamefont{Honma et~al.}(2008)}]{honma08}
\bibinfo{author}{\bibfnamefont{M.}~\bibnamefont{Honma}} \bibnamefont{et~al.},
  \bibinfo{journal}{RIKEN Accel. Prog. Rep.} \textbf{\bibinfo{volume}{41}},
  \bibinfo{pages}{32} (\bibinfo{year}{2008}).

\bibitem[{\citenamefont{Kahana et~al.}(1969)\citenamefont{Kahana, Lee, and
  Scott}}]{kahana69}
\bibinfo{author}{\bibfnamefont{S.}~\bibnamefont{Kahana}},
  \bibinfo{author}{\bibfnamefont{H.~C.} \bibnamefont{Lee}}, \bibnamefont{and}
  \bibinfo{author}{\bibfnamefont{C.~K.} \bibnamefont{Scott}},
  \bibinfo{journal}{Phys. Rev.} \textbf{\bibinfo{volume}{180}},
  \bibinfo{pages}{956} (\bibinfo{year}{1969}).

\bibitem[{\citenamefont{Dufour and Zuker}(1996)}]{dufour96}
\bibinfo{author}{\bibfnamefont{M.}~\bibnamefont{Dufour}} \bibnamefont{and}
  \bibinfo{author}{\bibfnamefont{A.~P.} \bibnamefont{Zuker}},
  \bibinfo{journal}{Phys. Rev. C} \textbf{\bibinfo{volume}{54}},
  \bibinfo{pages}{1641} (\bibinfo{year}{1996}).

\bibitem[{\citenamefont{Thompson}(1988)}]{thompson88}
\bibinfo{author}{\bibfnamefont{I.}~\bibnamefont{Thompson}},
  \bibinfo{journal}{Comp. Phys. Rep.} \textbf{\bibinfo{volume}{7}},
  \bibinfo{pages}{167} (\bibinfo{year}{1988}).

\bibitem[{\citenamefont{Perey and Perey}(1976)}]{perey76}
\bibinfo{author}{\bibfnamefont{C.~M.} \bibnamefont{Perey}} \bibnamefont{and}
  \bibinfo{author}{\bibfnamefont{F.~G.} \bibnamefont{Perey}},
  \bibinfo{journal}{Atomic Data Nucl. Data Tables}
  \textbf{\bibinfo{volume}{17}}, \bibinfo{pages}{1} (\bibinfo{year}{1976}).

\bibitem[{\citenamefont{Li et~al.}(2007)\citenamefont{Li, Liang, and
  Cai}}]{li07}
\bibinfo{author}{\bibfnamefont{X.}~\bibnamefont{Li}},
  \bibinfo{author}{\bibfnamefont{C.}~\bibnamefont{Liang}}, \bibnamefont{and}
  \bibinfo{author}{\bibfnamefont{C.}~\bibnamefont{Cai}},
  \bibinfo{journal}{Nuclear Physics A} \textbf{\bibinfo{volume}{789}},
  \bibinfo{pages}{103} (\bibinfo{year}{2007}).

\bibitem[{\citenamefont{Pang et~al.}(2009)\citenamefont{Pang, Roussel-Chomaz,
  Savajols, Varner, and Wolski}}]{pang09}
\bibinfo{author}{\bibfnamefont{D.~Y.} \bibnamefont{Pang}},
  \bibinfo{author}{\bibfnamefont{P.}~\bibnamefont{Roussel-Chomaz}},
  \bibinfo{author}{\bibfnamefont{H.}~\bibnamefont{Savajols}},
  \bibinfo{author}{\bibfnamefont{R.~L.} \bibnamefont{Varner}},
  \bibnamefont{and} \bibinfo{author}{\bibfnamefont{R.}~\bibnamefont{Wolski}},
  \bibinfo{journal}{Phys. Rev. C} \textbf{\bibinfo{volume}{79}},
  \bibinfo{pages}{024615} (\bibinfo{year}{2009}).

\bibitem[{\citenamefont{Daehnick et~al.}(1980)\citenamefont{Daehnick, Childs,
  and Vrcelj}}]{daehnick80}
\bibinfo{author}{\bibfnamefont{W.~W.} \bibnamefont{Daehnick}},
  \bibinfo{author}{\bibfnamefont{J.~D.} \bibnamefont{Childs}},
  \bibnamefont{and} \bibinfo{author}{\bibfnamefont{Z.}~\bibnamefont{Vrcelj}},
  \bibinfo{journal}{Phys. Rev. C} \textbf{\bibinfo{volume}{21}},
  \bibinfo{pages}{2253} (\bibinfo{year}{1980}).

\bibitem[{\citenamefont{Becchetti and Greenlees}(1969)}]{becchetti69}
\bibinfo{author}{\bibfnamefont{F.~D.} \bibnamefont{Becchetti}}
  \bibnamefont{and} \bibinfo{author}{\bibfnamefont{G.~W.}
  \bibnamefont{Greenlees}}, \bibinfo{journal}{Phys. Rev.}
  \textbf{\bibinfo{volume}{182}}, \bibinfo{pages}{1190} (\bibinfo{year}{1969}).

\bibitem[{\citenamefont{Varner et~al.}(1991)\citenamefont{Varner, Thompson,
  McAbee, Ludwig, and Clegg}}]{varner97}
\bibinfo{author}{\bibfnamefont{R.}~\bibnamefont{Varner}},
  \bibinfo{author}{\bibfnamefont{W.~J.} \bibnamefont{Thompson}},
  \bibinfo{author}{\bibfnamefont{T.~L.} \bibnamefont{McAbee}},
  \bibinfo{author}{\bibfnamefont{E.~J.} \bibnamefont{Ludwig}},
  \bibnamefont{and} \bibinfo{author}{\bibfnamefont{T.}~\bibnamefont{Clegg}},
  \bibinfo{journal}{Physics Reports} \textbf{\bibinfo{volume}{201}},
  \bibinfo{pages}{57 } (\bibinfo{year}{1991}).

\bibitem[{\citenamefont{Thompson}(2013)}]{thompson12}
\bibinfo{author}{\bibfnamefont{I.}~\bibnamefont{Thompson}}
  (\bibinfo{publisher}{Fifty Years of Nuclear BCS: Pairing in Finite Systems},
  \bibinfo{address}{Ricardo A Broglia, Vladimir Zelevinsky},
  \bibinfo{year}{2013}), \bibinfo{note}{http://arxiv.org/abs/1204.3054v2}.

\bibitem[{\citenamefont{Zieli\ifmmode~\acute{n}\else \'{n}\fi{}ska
  et~al.}(2009)}]{zielinska09}
\bibinfo{author}{\bibfnamefont{M.}~\bibnamefont{Zieli\ifmmode~\acute{n}\else
  \'{n}\fi{}ska}} \bibnamefont{et~al.}, \bibinfo{journal}{Phys. Rev. C}
  \textbf{\bibinfo{volume}{80}}, \bibinfo{pages}{014317}
  (\bibinfo{year}{2009}).

\end{thebibliography}

\end{document}